\documentclass[twocolumn,showpacs,preprintnumbers]{revtex4}
\usepackage{amsmath,amssymb}
\usepackage{graphicx}
\usepackage{dcolumn}
\usepackage{epsfig,color}
\usepackage{bm}
\usepackage{verbatim}

\begin{document}

\title{$\eta_{\rm c}$- and $J/\psi$-isoscalar meson bound states in the hadro-charmonium picture}
 
\author{J. Ferretti}
\affiliation{CAS Key Laboratory of Theoretical Physics, Institute of Theoretical Physics, Chinese Academy of Sciences, Beijing 100190, China}

\begin{abstract}
We study $\eta_{\rm c}$- and $J/\psi$-isoscalar meson bound states in the hadro-charmonium picture.
In the hadro-charmonium, the four $q\bar q c \bar c$ quarks are arranged in terms of a compact charm-anticharm pair, $c \bar c$, embedded in light hadronic matter, $q \bar q$, with $q = u$, $d$ or $s$. 
The interaction between the charmonium core and the light matter can be written in terms of the multipole expansion in QCD, with the leading term being the $E1$ interaction with chromo-electric field ${\bf E}^a$.
The spectrum of $\eta_{\rm c}$- and $J/\psi$-isoscalar meson bound states is calculated and the results compared with the existing experimental data.
\end{abstract}
\maketitle

\section{Introduction}
Recent discoveries by Belle and BESIII Collaborations of charged and neutral exotic quarkonium-like resonances, which do not fit into a traditional quark-antiquark interpretation, have driven new interest in theoretical and experimental searches for exotics.
Charged states, like $Z_{\rm c}(3900)$ \cite{Ablikim,Liu}, $Z_{\rm c}(4025)$ \cite{Ablikim2}, $Z_{\rm b}(10610)$ and $Z_{\rm b}(10650)$ \cite{Bondar}, have similar features and must be made up of four valence quarks because of their exotic quantum numbers.  
There are also several examples of neutral exotic quarkonium-like resonances, the so-called $X$ states, whose unusual properties do not fit into a quark-antiquark classification \cite{Nakamura:2010zzi}. 

A famous example is the $X(3872)$ \cite{Choi:2003ue,Aaij:2013zoa}, whose quark structure is still an open puzzle.
This resonance is characterized by $J^{PC} = 1^{++}$ quantum numbers, a very narrow width, and a mass $50-100$ MeV lower than quark model (QM) predictions \cite{Nakamura:2010zzi}. 
The charmonium interpretation of the $X(3872)$ as a $\chi_{c1}(2^3P_1)$ state is incompatible with the present experimental data, because the difference between the calculated \cite{Godfrey:1985xj,Eichten:1978tg,Barnes:2005pb} and experimental \cite{Nakamura:2010zzi} values of the meson mass is larger than the typical error of a QM calculation, of the order of $30-50$ MeV.
Because of these discrepancies between theory and data, several alternative interpretations for $X$ states have been proposed in addition to quarkonium, including: I) Meson-meson molecules \cite{Tornqvist:1993ng,Swanson:2003tb,Hanhart:2007yq,Thomas:2008ja,Baru:2011rs,Valderrama:2012jv,Guo:2013sya}; II) The result of kinematic or threshold effects caused by virtual particles \cite{Heikkila:1983wd,Pennington:2007xr,Li:2009ad,Danilkin:2010cc,Ferretti:2012zz,Ferretti:2013faa,Lu:2016mbb}; III) Compact tetraquark (or diquark-antidiquark) states \cite{Jaffe:1976ih,Barbour:1979qi,Weinstein:1983gd,SilvestreBrac:1993ss,Brink:1998as,Maiani:2004vq,Barnea:2006sd,Santopinto:2006my,Anwar:2017toa}; IV) Hadro-quarkonia (hadro-charmonia) \cite{Dubynskiy:2008mq,Guo:2008zg,Guo:2009id,Voloshin:2013dpa,Li:2013ssa,Wang:2013kra,Brambilla:2015rqa,Cleven:2015era,Alberti:2016dru,Panteleeva:2018ijz}; V) The rescattering effects arising by anomalous triangular singularities \cite{Guo:2014iya,Szczepaniak:2015eza,Liu:2015taa}. For a review, see Refs. \cite{Guo:2017jvc,Seth:2012,Esposito:2014rxa,Olsen:2017bmm}.
Here, we focus on the hadro-charmonium picture.

The hadro-charmonium is a tetraquark configuration, where a compact $c \bar c$ state ($\psi$) is embedded in light hadronic matter ($\mathcal X$) \cite{Dubynskiy:2008mq}.
The interaction between the two components, $\psi$ and $\mathcal X$, takes place via a QCD analog of the van der Waals force of molecular physics. 
It can be written in terms of the multipole expansion in QCD \cite{Gottfried:1977gp,Voloshin:1978hc,Peskin:1979va}, with the leading term being the $E1$ interaction with chromo-electric field ${\bf E}^a$.

The hadrocharmonium picture was motivated by the observation that several charmonium-like states are only found in specific charmonium-light hadron final states. 
Some examples include $X(4260)$, observed in the $J/\Psi\pi\pi$ channel \cite{Aubert:2005rm}, $Z_{\rm c}(4430)$ discovered in $\psi(2S) \pi$ \cite{Mizuk:2009da}, $X(4360)$ and $X(4660)$, observed in $\psi(2S) \pi \pi$ \cite{Wang:2007ea,Aubert:2007zz}.
The recent BESIII observation of similar cross sections for $J/\Psi \pi^+\pi^-$ and $h_{\rm c} \pi^+\pi^-$ at 4.26 and 4.36 GeV in $e^+e^-$ collisions \cite{Ablikim,Ablikim:2013wzq} stimulated Li and Voloshin to extend the hadrocharmonium model by including also heavy-quark spin-symmetry breaking. As a result, $X(4260)$ and $X(4360)$ were described as a mixture of two hadrocharmonia, $\left|\psi_1 \right\rangle \sim \left|1^{+-}\right\rangle_{c \bar c} \otimes \left| 0^{-+} \right\rangle_{q \bar q}$ and $\psi_3 \sim \left|1^{--}\right\rangle_{c \bar c} \otimes \left| 0^{++} \right\rangle_{q \bar q}$, with a large mixing angle, $\theta_{\rm mix} \simeq 40^\circ$ \cite{Li:2013ssa,Cleven:2015era}.
Recently, the hadro-charmonium model was also used to discuss the emergence of $\phi-\psi(2S)$ bound states, including the principal decay modes \cite{Panteleeva:2018ijz}.
According to the previous study, the $\phi-\psi(2S)$ bound state is a good candidate for a tetraquark with hidden charm and strangeness.
See also Refs. \cite{Guo:2008zg,Guo:2009id}, where the $Y(4660)$ is interpreted as a $\Psi(2S)-{\rm f}_0$ bound state, with spin partner $\eta_{\rm c}(2S)-{\rm f}_0$, and Ref. \cite{Voloshin:2013dpa}, where a hadro-charmonium assignment for the $Z_{\rm c}(3900)$ is discussed.
 
In the present manuscript, we calculate the spectrum of $\eta_{\rm c}$- and $J/\psi$-isoscalar meson bound states under the hadro-charmonium hypothesis.
The $q\bar q c \bar c$ masses are computed by solving the Schr\"odinger equation for the hadro-charmonium potential \cite{Dubynskiy:2008mq}. This is approximated as a finite well whose width and size can be expressed as a function of the chromo-electric polarizability, $\alpha_{\psi \psi}$, and light meson radius. The chromo-electric polarizability is estimated in the framework of the $1/N_{\rm c}$ expansion \cite{Peskin:1979va,Eides:2015dtr}.
Finally, the hadro-charmonium masses and quantum numbers are compared with the existing experimental data.
Some tentative assignments are also discussed.

The hypothesis of charmed and bottom pentaquarks as light baryon-quarkonium bound states will also be investigated \cite{Luke:1992tm,Eides:2015dtr,Kaidalov:1992hd,Brodsky:1989jd}.

\section{A mass formula for the hadro-charmonium}
The hadro-charmonium is a tetraquark configuration, where a compact $c \bar c$ state ($\psi$) is embedded in light hadronic matter ($\mathcal X$) \cite{Dubynskiy:2008mq}.
The interaction between the charmonium core, $\psi$, and the gluonic field inside the light-meson, $\mathcal X$, can be written in terms of the QCD multipole expansion \cite{Gottfried:1977gp,Voloshin:1978hc,Peskin:1979va}, considering as leading term the $E1$ interaction with chromo-electric field ${\bf E}^a$ \cite{Dubynskiy:2008mq,Kaidalov:1992hd}. 

\begin{figure}[htbp] 
\centering 
\includegraphics[width=4cm]{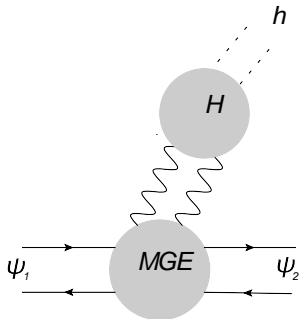}
\caption{Hidden-flavor amplitude $\psi_1 \rightarrow \psi_2 h$ in the QCD multipole expansion approach. Here, $\psi_1$ and $\psi_2$ are the initial and final charmonium states, $h$ light hadron(s). The two vertices are those of the multipole gluon emission ($MGE$) and hadronization ($H$).}
\label{fig:gluon-trans}
\end{figure}

The effective Hamiltonian we consider is the same describing a $\psi_2 \rightarrow \psi_1$ transition in the chromo-electric field. It can can be written as \cite{Voloshin:2007dx}
\begin{equation}
	\label{eqn:Heff}
	H_{\rm eff} = - \frac{1}{2} \alpha_{ij}^{(12)} E_i^a E_j^a  \mbox{ },
\end{equation}
where
\begin{equation}
	\label{eqn:alpha12}
	\alpha_{ij}^{(12)} = \frac{1}{16} \left\langle \psi_1 \right| \xi^a r_i \mathcal G r_j \xi^a \left| \psi_2 \right\rangle 
\end{equation}
is the chromo-electric polarizability. It is expressed in terms of the Green function $\mathcal G$ of the heavy-quark pair in a color octet state (having the same color quantum numbers as a gluon), the relative coordinate between the quark and the antiquark, $\bf r$, and the difference between the color generators acting on them, $\xi^a = t_1^a - t_2^a$. A schematic representation of a hidden-flavor $\psi_{\rm 1} \rightarrow \psi_{\rm 2} + h$ transition in the QCD multipole expansion approach is given in Fig. \ref{fig:gluon-trans}. 
%%%%%%%%%%%%%%%%%%%%%%%%%%%%%%%%%%%%%%%%
%%%%%%%%%%%%%%%%%%%%%%%%%%%%%%%%%%%%%%%%
Here, $\psi_1$ and $\psi_2$ are the initial and final charmonium states, $h$ light hadron(s). The two vertices are those of the multipole gluon emission ($MGE$) and hadronization ($H$).
\begin{comment}
In the particular case of a transition between $S$-wave states, Eq. (\ref{eqn:alpha12}) reduces to a scalar
\begin{equation}
	\label{eqn:alpha12-Swave}
	\alpha_{ij}^{(12)} = \alpha^{(12)} \delta_{ij} \mbox{ } {\bm \epsilon}_1 \cdot {\bm \epsilon}_2  \mbox{ },
\end{equation}
where ${\bm \epsilon}_{1,2}$ are the polarization amplitudes of $\psi_{1,2}$.
\end{comment}

In order to calculate the hadrocharmonium masses, we have to compute the expectation value of Eq. (\ref{eqn:alpha12}) on the charmonium state $\left| \psi \right\rangle$, i.e. the diagonal chromo-electric polarizability $\alpha_{\psi \psi}$, and also the diagonal matrix elements $\left\langle \mathcal X \right| E_i^a E_i^a \left| \mathcal X \right\rangle$.

\subsection{Diagonal chromo-electric polarizability}
\label{Diagonal chromo-electric polarizability}
In the following, we discuss three possible prescriptions for the diagonal chromo-electric polarizabilities, $\alpha_{\psi \psi}$.
\begin{enumerate}
\item It is possible to provide an estimation of the off-diagonal chromo-electric polarizability, $\alpha_{\psi \psi'}$, from the decay rate $\psi(2S) \rightarrow J/\psi \pi^+ \pi^-$; the resulting value is \cite{Voloshin:2007dx,Sibirtsev:2005ex}:
\begin{equation}
	\label{eqn:alpha.eq.2}
	\alpha_{\psi \psi'} \approx 2 \mbox{ GeV}^{-3}  \mbox{ }.
\end{equation} 
After introducing final state interactions, $\alpha_{\psi \psi'}$ from Eq. (\ref{eqn:alpha.eq.2}) is reduced to about $\frac{1}{3}$ of its value \cite{Guo:2006ya}.
Even if we expect diagonal $\alpha$ parameters, $\alpha_{\psi \psi}$, to be larger than off-diagonal ones, $\alpha_{\psi \psi'}$, one possibility is to take $\alpha_{\psi \psi} = \alpha_{\psi \psi'} = 2 \mbox{ GeV}^{-3}$. 
Because of the smallness of (\ref{eqn:alpha.eq.2}), this prescription only gives rise to a few weakly-bounded states, like $\eta_{\rm c}(2S) \otimes {\rm f}_0'$ and $\psi(2S) \otimes {\rm f}_0'$, with masses of 4981 and 5027 MeV, respectively. Thus, this first possibility is neglected.
\item Alternately, one can calculate the chromo-electric polarizability by considering quarkonia as pure Coulombic systems.
While this is a very good approximation in the case of $b \bar b$ states, one may object that it is questionable in the case of charmonia.

The perturbative result in the framework of the $1/N_{\rm c}$ expansion is \cite{Peskin:1979va,Eides:2015dtr}
\begin{equation}
	\label{alphacoulomb}
	\alpha_{\psi \psi}(nS)=\frac{16\pi n^2 c_n a_0^3}{3g_{\rm c}^2 N_{\rm c}^2} \mbox{ }.
\end{equation}
Here, $n$ is the radial quantum number; $c_1=\frac{7}{4}$ and $c_2=\frac{251}{8}$; $N_{\rm c} = 3$ is the number of colors; $g_{\rm c} = \sqrt{4\pi \alpha_{\rm s}} \simeq 2.5$, with $\alpha_{\rm s}$ being the QCD running coupling constant; finally, 
\begin{equation}
	\label{eqn:a0-Bohr}
	a_0 = \frac{2}{m_{\rm c} C_{\rm F} \alpha_{\rm s}}
\end{equation}
is the Bohr radius of nonrelativistic charmonium \cite{Brambilla:2015rqa}, with $C_{\rm F} = \frac{N_{\rm c}^2 - 1}{2 N_{\rm c}}$ and $m_{\rm c} = 1.5$ GeV.
By using Eqs. (\ref{alphacoulomb}) and (\ref{eqn:a0-Bohr}) and the previous values of the constants and parameters, one obtains 
\begin{subequations}
\begin{equation}
	\alpha_{\psi \psi}(1S) \simeq 4.1 \mbox{ GeV}^{-3}
\end{equation}	 
and 
\begin{equation}
	\alpha_{\psi \psi}(2S) \simeq 296 \mbox{ GeV}^{-3}  \mbox{ }.
\end{equation}
\end{subequations}
As discussed in the following, the value of $\alpha_{\psi \psi}(1S)$ gives rise to hadro-charmonium states with binding energies $\mathcal{O}(10-100)$ MeV. 
On the contrary, the largeness of $\alpha_{\psi \psi}(2S)$ gives rise to unphysical states, characterized by negative masses.
A possible explanation is the following: $2S$ are larger than $1S$ $c \bar c$ states; thus, the QCD multipole expansion, where one assumes the quarkonium size to be much smaller than the soft-gluon wave-length, is not applicable anymore.

In the bottomonium case, considering $\alpha_{\rm s} \simeq 0.35$ and $m_{\rm b} \simeq 5.0$ GeV \cite{Brambilla:2015rqa}, one gets: $\alpha_{\Upsilon \Upsilon}(1S) \simeq 0.47$ GeV$^{-3}$ and $\alpha_{\Upsilon \Upsilon}(2S) \simeq 33$ GeV$^{-3}$.
$\alpha_{\Upsilon \Upsilon}(1S)$ may be too small to generate bounded states; on the contrary, $\alpha_{\Upsilon \Upsilon}(2S)$ may give rise to hadro-bottomonia with large binding energies, $\mathcal{O}(1)$ GeV, which may be unphysical.

\item The third possibility is to calculate the expectation value of Eq. (\ref{eqn:alpha12}) on charmonia by inserting string-vibrational or continuum-octet intermediate states \cite{Yan:1980,Yan:1981,Yan:1988,Li:1986,Zhou:1991,Brambilla:2015rqa} in the matrix element of Eq. (\ref{eqn:alpha12}). 

Specifically, Eq. (\ref{eqn:alpha12}) can be re-written as \cite{Voloshin:1978hc,Voloshin:1980zf,Moxhay:1987ch}
\begin{equation}
	\label{eqn:pert3}
	\alpha_{\psi \psi} = \frac{1}{24} \left\langle \psi \right| r_i G_8 r_i \left| \psi \right\rangle \mbox{ }. 
\end{equation}
Here, the condition $\left\langle \mbox{singlet} \right| \xi^a \xi^b \left| \mbox{singlet} \right\rangle = \frac{2}{3} \delta^{ab}$ is used, because the operator $\xi^a$ turns a singlet state into an octet one, and vice-versa (only the octet states contribute), and 
\begin{equation}
	\label{eqn:G8}
	G_8 = \frac{1}{E_\psi - E_8} = \sum_{k\ell} \frac{\left|\nu k \ell \right\rangle \left\langle \nu k \ell \right|}{E_\psi - E^\nu_{k\ell}}
\end{equation}	
is the color-octet Green's function. Here, $E_\psi$ and $E^\nu_{k\ell}$ are charmonium and string-vibrational state \cite{Tye,Isgur-Paton} energies.
After introducing the propagator of Eq. (\ref{eqn:G8}) in (\ref{eqn:pert3}), the chromo-electric polarizability calculation essentially reduces to evaluating dipole matrix elements between quarkonium and string-vibrational states. 
\end{enumerate}

\subsection{$E_i^a E_i^a$ product}
The product $E_i^a E_i^a$ in Eq. (\ref{eqn:Heff}) can be re-written using the anomaly in the trace of the energy-momentum tensor $\theta_{\mu \nu}$ in QCD \cite{Voloshin:1980zf},
\begin{equation}
  \label{eqn:thetamumu}
	\begin{array}{rcl}
	\theta_\mu^\mu & = & - \frac{9}{32 \pi^2} \mbox{ } G_{\mu\nu}^a G^{a\mu\nu}  \\ 
	& = & \frac{9}{16 \pi^2} \mbox{ } (E_i^a E_i^a - B_i^a B_i^a)
	\end{array} \mbox{ }, 
\end{equation}
where $B_i^a$ is the chromo-magnetic field.
If we neglect the contribution due to the chromo-magnetic fields, which is expected to be smaller than the chromo-electric one \cite{Voloshin:2007dx}, Eq. (\ref{eqn:thetamumu}) can be re-written as:
\begin{equation}
  \label{eqn:thetamumu2}
	E_i^a E_i^a \approx \frac{16 \pi^2}{9} \mbox{ } \theta_\mu^\mu  \mbox{ }. 
\end{equation}
The expectation value of the operator $\theta_\mu^\mu$ on a generic state $\mathcal X$ is given by \cite{Dubynskiy:2008mq}
\begin{equation}
	\left\langle \mathcal X \right| \theta_\mu^\mu ({\bf q} = 0) \left| \mathcal X \right\rangle = M_{\mathcal X}  \mbox{ },
\end{equation}
where a non-relativistic normalization for $\mathcal X$, $\left\langle \mathcal X | \mathcal X \right\rangle = 1$, is assumed.

\subsection{An Hamiltonian for the hadro-charmonium}
The effective potential $V_{\rm hc}$, describing the coupling between $\psi$ and $\mathcal X$, can be approximated as a finite well \cite{Dubynskiy:2008mq}
\begin{equation}
	\label{eqn:Vd3r}
	\int_0^{R_{\mathcal X}} d^3r \mbox{ } V_{hc} \approx - \frac{8 \pi^2}{9} \mbox{ } \alpha_{\psi \psi} M_{\mathcal X}  \mbox{ },
\end{equation}
where
\begin{equation}
	\label{eqn:radius}
	R_{\mathcal X} = \int_0^{\infty} d^3r \Psi^*_{\mathcal X}({\bf r}) r \Psi_{\mathcal X}({\bf r})
\end{equation}
is the radius of the light meson $\mathcal X$ \cite{Godfrey:1985xj}. 
Thus, we have:
\begin{equation}
\label{eqn:Vhc}
	V_{\rm hc}(r) = \left\{ \begin{array}{ccc} -\frac{2\pi\alpha_{\psi \psi}M_{\mathcal X}}{3R_{\mathcal X}^3} & \mbox{for} & r < R_{\mathcal X} \\
	                         0   & \mbox{for} & r > R_{\mathcal X}  
		                \end{array}  \right.  \mbox{ }.
\end{equation}
By analogy with calculations of the interaction between heavy quarkonia and the nuclear medium \cite{Voloshin:2007dx,Sibirtsev:2005ex,Kaidalov:1992hd}, we get a potential that is a constant square well inside the light meson $\mathcal X$ and null outside.
We can estimate the order of magnitude of the strength of $V_{\rm hc}$ by introducing into Eq. (\ref{eqn:Vhc}) typical values for $R_{\mathcal X}$ and $M_{\mathcal X}$. If we take $R_{\mathcal X} = 0.5$ fm, $M_{\mathcal X} = 1$ GeV and $\alpha_{\psi \psi}$ from Eq. (\ref{eqn:alpha.eq.2}), we get a potential well with a depth of the order of 250 MeV.
The Hamiltonian of the hadro-charmonium system also contains a kinetic energy term, 
\begin{equation}
	\label{eqn:Thc}
	T_{\rm hc} = \frac{k^2}{2 \mu}  \mbox{ },
\end{equation}
where ${\bf k}$ is the relative momentum (with conjugate coordinate ${\bf r}$) between $\psi$ and $\mathcal X$, and $\mu$ the reduced mass of the $\psi \mathcal X$ system.

The total hadro-charmonium Hamiltonian is thus:
\begin{equation}
	\label{eqn:Hhc}
	H_{\rm hc} = M_\psi + M_{\mathcal X} + V_{\rm hc}(r) + T_{\rm hc}  \mbox{ }.
\end{equation}

%%%%%%%%%%%%%%%%%%%%%%%%%%
\begin{table*}
\begin{tabular}{cccccc} 
\hline 
\hline
Composition & Quark content & $J_{\rm hc}^{PC}$ & Binding [MeV] & Mass [MeV] & Assignment \\
\hline
$\eta_{\rm c} \otimes \eta'$ & $c \bar c s \bar s$ & $0^{++}$ & 12 & 3929 & $X(3915)$ \\
$\eta_{\rm c} \otimes {\rm f}_0$ & $c \bar c q \bar q$ & $0^{-+}$, $1^{-+}$, $2^{-+}$ & 28 & 3946 & $X(3940)$ \\
$\eta_{\rm c} \otimes \phi$ & $c \bar c s \bar s$ & $1^{+-}$ & 20 & 3983 & -- \\
$J/\psi \otimes \eta'$ & $c \bar c s \bar s$ & $1^{+-}$ & 13 & 4042 & -- \\
$J/\psi \otimes {\rm f}_0$ & $c \bar c q \bar q$ & $0^{-+}$, $1^{--}$, $1^{-+}$, $2^{-+}$, $2^{--}$, $3^{--}$ & 29 & 4058 & -- \\
$J/\psi \otimes \phi$ & $c \bar c s \bar s$ & $0^{++}$, $1^{+-}$, $2^{++}$ & 21 & 4096 & -- \\
$\eta_{\rm c} \otimes {\rm h}_1$ & $c \bar c q \bar q$ & $1^{--}$ & 37 & 4116 & -- \\
$\eta_{\rm c} \otimes {\rm f}_0'$  & $c \bar c s \bar s$ & $0^{-+}$, $1^{-+}$, $2^{-+}$ & 151 & 4191 & -- \\
$\eta_{\rm c} \otimes {\rm f}_1$ & $c \bar c q \bar q$ & $0^{-+}$, $1^{-+}$, $2^{-+}$ & 61 & 4204 & $X(4160)$ \\
$J/\psi \otimes {\rm h}_1$ & $c \bar c q \bar q$ & $0^{-+}$, $1^{-+}$, $2^{-+}$ & 38 & 4229 & -- \\
$\eta_{\rm c} \otimes {\rm f}_2$ & $c \bar c q \bar q$ & $0^{-+}$, $1^{-+}$, $2^{-+}$ & 25 & 4234 & -- \\
$\eta_{\rm c} \otimes {\rm h}_1'$ & $c \bar c s \bar s$ & $1^{--}$ & 105 & 4285 & -- \\
$\eta_{\rm c} \otimes {\rm f}_1'$ & $c \bar c s \bar s$ & $0^{-+}$, $1^{-+}$, $2^{-+}$ & 118 & 4292 & -- \\
$J/\psi \otimes {\rm f}_0'$ & $c \bar c s \bar s$ & $0^{-+}$, $1^{--}$, $1^{-+}$, $2^{-+}$, $2^{--}$, $3^{--}$ & 153 & 4303 & -- \\
$J/\psi \otimes {\rm f}_1$ & $c \bar c q \bar q$ & $0^{-+}$, $1^{--}$, $1^{-+}$, $2^{-+}$, $2^{--}$, $3^{--}$ & 62 & 4317 & $Y(4260)$ \\
$J/\psi \otimes {\rm f}_2$ & $c \bar c q \bar q$ & $0^{-+}$, $1^{--}$, $1^{-+}$, $2^{-+}$, $2^{--}$, $3^{--}$ & 26 & 4346 & $Y(4360)$ \\
$J/\psi \otimes {\rm h}_1'$ & $c \bar c s \bar s$ & $0^{-+}$, $1^{-+}$, $2^{-+}$ & 107 & 4397 & -- \\
$J/\psi \otimes {\rm f}_1'$ & $c \bar c s \bar s$ & $0^{-+}$, $1^{--}$, $1^{-+}$, $2^{-+}$, $2^{--}$, $3^{--}$ & 120 & 4404 & -- \\
$\eta_{\rm c} \otimes {\rm f}_2'$ & $c \bar c s \bar s$ & $0^{-+}$, $1^{-+}$, $2^{-+}$ & 85 & 4423 & -- \\
$J/\psi \otimes {\rm f}_2'$  & $c \bar c s \bar s$ & $0^{-+}$, $1^{--}$, $1^{-+}$, $2^{-+}$, $2^{--}$, $3^{--}$ & 87 & 4535 & -- \\
\hline 
\hline
\end{tabular}
\caption{Hadro-charmonium model predictions (fourth and fifth columns), calculated by solving the Schr\"odinger equation (\ref{eqn:Hhc}) with the chromo-electric polarizability of Eq. (\ref{alphacoulomb}).
The ${\rm f}_0'$ mass used in the calculations, $M_{{\rm f}_0'} = 1359$ MeV, is calculated in the relativized quark model \cite{Godfrey:1985xj}.} 
\label{tab:hadro-charmonium-spectrum}  
\end{table*}
%%%%%%%%%%%%%%%%%%%%%%%%%%

\section{Results and discussion}
Below, we calculate the spectrum of $\eta_{\rm c}$- and $J/\psi$-isoscalar meson bound states in the hadro-charmonium picture by solving the eigenvalue problem of Eq. (\ref{eqn:Hhc}).
The time-independent Schr\"odinger equation is solved numerically by means of both Multhopp method, see \cite[Sec. 2.4]{Richard} and \cite[Sec. II.D]{Basdevant}, and a finite differences algorithm \cite[Vol. 3, Sec. 16-6]{Feynman-Lectures} as a check. 
The theoretical predictions are extracted by using the prescription 2. for the chromo-electric polarizability of Sec. \ref{Diagonal chromo-electric polarizability}.

The calculated hadro-charmonium spectrum is shown in Table \ref{tab:hadro-charmonium-spectrum}; here, we also try some tentative assignments to experimental $X$ states. See \cite[Table I]{Olsen:2017bmm}.

The hadro-charmonium quantum numbers are shown in the third column of Table \ref{tab:hadro-charmonium-spectrum}. They are obtained by combining those of the charmonium core, $\psi$, and light meson, $\mathcal X$, as
\begin{equation}
	\left| \Phi_{\rm hc} \right\rangle = \left| (L_\psi, L_{\mathcal X}) L_{\rm hc}, (S_\psi, S_{\mathcal X}) S_{\rm hc}; J_{\rm hc}^{PC} \right\rangle  \mbox{ },
\end{equation}
where the hadro-charmonium $P$- and $C$-parity are given by: $P=(-1)^{L_{\rm hc}}$ and $C = (-1)^{L_{\rm hc} + S_{\rm hc}}$.

Starting from the lowest part of the spectrum, the $X(3915)$, observed by Belle and BaBar in $B \rightarrow K + (J\psi \omega)$ \cite{delAmoSanchez:2010jr} and $e^+ e^- \rightarrow e^+ e^- + (J\psi \omega)$ \cite{Uehara:2009tx}, is interpreted as a $\eta_{\rm c} \otimes \eta'$ hadro-charmonium state.
The $X(3940)$, discovered by Belle in $e^+ e^- \rightarrow J/\Psi + \mbox{ anything}$ \cite{Abe:2007jna} and later observed in $e^+ e^- \rightarrow J/\Psi + (D^* \bar D)$ \cite{Abe:2007sya}, is here interpreted as a $\eta_{\rm c} \otimes {\rm f}_0$ state.
The $X(4160)$, observed by Belle in $e^+ e^- \rightarrow J/\Psi + (D^*\bar D^*)$ \cite{Abe:2007sya}, may be interpreted as a $\eta_{\rm c} \otimes {\rm f}_1$ state with $0^{-+}$ quantum numbers.
The $X(4260)$, observed by BaBar \cite{Aubert:2005rm,Lees:2012cn}, CLEO \cite{He:2006kg} and Belle \cite{Yuan:2007sj,Liu} in $e^+ e^- \rightarrow \gamma + (J/\Psi \pi^+ \pi^-)$, and $X(4360)$, observed by BaBar \cite{Aubert:2007zz,Lees:2012pv} and Belle \cite{Wang:2007ea,Wang:2014hta} in $e^+ e^- \rightarrow \gamma + [\Psi(2S) \pi^+ \pi^-]$, and BESIII \cite{Ablikim:2016qzw} in $J/\Psi \pi^+\pi^-$ and $h_{\rm c} \pi^+\pi^-$, are both characterized by $1^{--}$ quantum numbers.
According to our results, $X(4260)$ and $X(4360)$ may be described in terms of $J/\psi \otimes {\rm f}_1$ and $J/\psi \otimes {\rm f}_2$ states, respectively.
In Refs. \cite{Li:2013ssa,Wang:2013kra,Cleven:2015era}, they are interpreted as a mixture of two hadrocharmonia, $\left|\psi_1 \right\rangle \sim \left|1^{+-}\right\rangle_{c \bar c} \otimes \left| 0^{-+} \right\rangle_{q \bar q}$ and $\psi_3 \sim \left|1^{--}\right\rangle_{c \bar c} \otimes \left| 0^{++} \right\rangle_{q \bar q}$, with a large mixing angle, $\theta_{\rm mix} \simeq 40^\circ$. The mixing is due to the exchange of one chromo-electric and one chromo-magnetic gluon between the hadro-charmonium $c \bar c$ cores.

Finally, it is worth noticing that: I) The quantum number assignments in Table \ref{tab:hadro-charmonium-spectrum} for several states are not univocal. A possible way to distinguish between them is to calculate the hadro-charmonium main decay amplitudes and compare the theoretical results with the data; II) The results strongly depend on the chromo-electric polarizability, $\alpha_{\psi \psi}$. Up to now, the value of $\alpha_{\psi \psi}$ cannot be fitted to the experimental data; it has to be estimated phenomenologically.
Because of this, it represents one of the main sources of theoretical uncertainty on the results; III) In the calculation of $\left\langle \mathcal X \right| \theta_\mu^\mu ({\bf q} = 0) \left| \mathcal X \right\rangle$ matrix elements on light mesons, $\mathcal X$, the contributions due to the chromo-magnetic field, ${\bf B}^a$, are neglected. This may represent another source of theoretical uncertainties; IV) By combining $\psi$ and $\mathcal X$ quantum numbers, several $J_{\rm hc}^{PC}$ configurations are obtained. Thus, once the value of the $J/\psi$ and $\eta_{\rm c}$ chromo-electric polarizability is measured (and thus the main source of theoretical uncertainties removed), it would be interesting to introduce spin-orbit and spin-spin corrections in order to split the degenerate configurations.

\begin{acknowledgments}
The author is grateful to Prof. Feng-Kun Guo for useful discussions and suggestions. 
\end{acknowledgments}


\begin{thebibliography}{}

\bibitem{Ablikim}
  M. Ablikim {\it et al.} [BESIII Collaboration],
  Phys.\ Rev.\ Lett. {\bf 110}, 252001 (2013).

\bibitem{Liu}
  Z. Q. Liu {\it et al.} [Belle Collaboration],
  Phys.\ Rev.\ Lett. {\bf 110}, 252002 (2013).

\bibitem{Ablikim2}
  M.~Ablikim {\it et al.} [BESIII Collaboration],
  %``Observation of a Charged Charmoniumlike Structure $Z_c$(4020) and Search for the $Z_c$(3900) in $e^+e^- \to ?^+?^-h_c$,''
  Phys.\ Rev.\ Lett.\  {\bf 111}, no. 24, 242001 (2013);
  %``Observation of a charged charmoniumlike structure in $e^+e^- \to (D^{*} \bar{D}^{*})^{\pm} \pi^\mp$ at $\sqrt{s}=4.26$GeV,''
  {\bf 112}, no. 13, 132001 (2014).
	
\bibitem{Bondar}
  A. Bondar {\it et al.} [Belle Collaboration],
  Phys.\ Rev.\ Lett. {\bf 108}, 122001 (2012).	
  
\bibitem{Nakamura:2010zzi}
  C. Patrignani {\it et al.} (Particle Data Group), 
  Chinese Physics C {\bf 40}, 100001 (2016). 	
	
\bibitem{Choi:2003ue}
  S.~K.~Choi {\it et al.} [Belle Collaboration],
  %``Observation of a narrow charmonium - like state in exclusive B+- ---> K+- pi+ pi- J / psi decays,''
  Phys.\ Rev.\ Lett.\  {\bf 91}, 262001 (2003).
  
\bibitem{Aaij:2013zoa}
  R.~Aaij {\it et al.}  [LHCb Collaboration],
  %``Determination of the X(3872) meson quantum numbers,''
  Phys.\ Rev.\ Lett.\  {\bf 110}, 222001 (2013).	
  
\bibitem{Eichten:1978tg}
  E.~Eichten, K.~Gottfried, T.~Kinoshita, K.~D.~Lane and T.~-M.~Yan,
  %``Charmonium: The Model,''
  Phys.\ Rev.\ D {\bf 17} (1978) 3090
  [Erratum-ibid.\ D {\bf 21} (1980) 313];
  %``Charmonium: Comparison with Experiment,''
  Phys.\ Rev.\ D {\bf 21} (1980) 203.  
	
\bibitem{Godfrey:1985xj}
  S.~Godfrey and N.~Isgur,
  %``Mesons in a Relativized Quark Model with Chromodynamics,''
  Phys.\ Rev.\ D {\bf 32}, 189 (1985).
  
\bibitem{Barnes:2005pb}
  T.~Barnes, S.~Godfrey and E.~S.~Swanson,
  %``Higher charmonia,''
  Phys.\ Rev.\ D {\bf 72}, 054026 (2005).       
  		
\bibitem{Tornqvist:1993ng} 
  N.~A.~T\"ornqvist,
  %``From the deuteron to deusons, an analysis of deuteron - like meson meson bound states,''
  Z.\ Phys.\ C {\bf 61}, 525 (1994);
  %``Isospin breaking of the narrow charmonium state of Belle at 3872-MeV as a deuson,''
  Phys.\ Lett.\ B {\bf 590}, 209 (2004). 
  
\bibitem{Swanson:2003tb}
  E.~S.~Swanson,
  %``Short range structure in the X(3872),''
  Phys.\ Lett.\ B {\bf 588}, 189 (2004);
  %``Diagnostic decays of the X(3872),''
  {\bf 598}, 197 (2004).

\bibitem{Hanhart:2007yq}
  C.~Hanhart, Y.~S.~Kalashnikova, A.~E.~Kudryavtsev and A.~V.~Nefediev,
  %``Reconciling the X(3872) with the near-threshold enhancement in the D0 anti-D*0 final state,''
  Phys.\ Rev.\ D {\bf 76}, 034007 (2007).
  
\bibitem{Thomas:2008ja} 
  C.~E.~Thomas and F.~E.~Close,
  %``Is X(3872) a molecule?,''
  Phys.\ Rev.\ D {\bf 78}, 034007 (2008).
  
\bibitem{Guo:2013sya} 
  M.~Cleven, F.~K.~Guo, C.~Hanhart and U.~G.~Meissner,
  %``Bound state nature of the exotic $Z_b$ states,''
  Eur.\ Phys.\ J.\ A {\bf 47}, 120 (2011);
  F.~K.~Guo, C.~Hidalgo-Duque, J.~Nieves and M.~P.~Valderrama,
  %``Consequences of Heavy Quark Symmetries for Hadronic Molecules,''
  Phys.\ Rev.\ D {\bf 88}, 054007 (2013).     
  
\bibitem{Baru:2011rs}
  V.~Baru {\it et al.},
  %``Three-body $D\bar{D}\pi$ dynamics for the X(3872),''
  Phys.\ Rev.\ D {\bf 84}, 074029 (2011).    
  
\bibitem{Valderrama:2012jv} 
  M.~P.~Valderrama,
  %``Power Counting and Perturbative One Pion Exchange in Heavy Meson Molecules,''
  Phys.\ Rev.\ D {\bf 85}, 114037 (2012).  

\bibitem{Heikkila:1983wd} 
  K.~Heikkila, S.~Ono and N.~A.~Tornqvist,
  %``HEAVY c anti-c AND b anti-b QUARKONIUM STATES AND UNITARITY EFFECTS,''
  Phys.\ Rev.\ D {\bf 29}, 110 (1984)
  Erratum: [Phys.\ Rev.\ D {\bf 29}, 2136 (1984)].  
 		
\bibitem{Pennington:2007xr}
  M.~R.~Pennington and D.~J.~Wilson,
  %``Decay channels and charmonium mass-shifts,''
  Phys.\ Rev.\ D {\bf 76}, 077502 (2007).    
  
\bibitem{Li:2009ad}
  B.~-Q.~Li, C.~Meng and K.~-T.~Chao,
  %``Coupled-Channel and Screening Effects in Charmonium Spectrum,''
  Phys.\ Rev.\ D {\bf 80}, 014012 (2009).  

\bibitem{Danilkin:2010cc}
  I.~V.~Danilkin and Y.~A.~Simonov,
  %``Dynamical origin and the pole structure of X(3872),''
  Phys.\ Rev.\ Lett.\  {\bf 105}, 102002 (2010). 
  
\bibitem{Ferretti:2012zz} 
  J.~Ferretti, G.~Galat\`a, E.~Santopinto and A.~Vassallo,
  %``Bottomonium self-energies due to the coupling to the meson-meson continuum,''
  Phys.\ Rev.\ C {\bf 86}, 015204 (2012);  
  J.~Ferretti, G.~Galat\`a and E.~Santopinto, 
  Phys.\ Rev.\ D {\bf 90}, 054010 (2014);	        
  J.~Ferretti and E.~Santopinto,
  %``Higher mass bottomonia,''
  Phys.\ Rev.\ D {\bf 90}, 094022 (2014);
  %``Threshold effects in $\chi_{\rm c}(2P)$ and $\chi_{\rm b}(3P)$ multiplets and $J/\Psi \rho, J/\Psi \omega$ hidden-flavor strong decays of the $X(3872)$,''
  arXiv:1806.02489.  
  
\bibitem{Ferretti:2013faa} 
  J.~Ferretti, G.~Galat\`a and E.~Santopinto,
  %``Interpretation of the X(3872) as a charmonium state plus an extra component due to the coupling to the meson-meson continuum,''
  Phys.\ Rev.\ C {\bf 88}, no. 1, 015207 (2013).  
  
\bibitem{Lu:2016mbb} 
  Y.~Lu, M.~N.~Anwar and B.~S.~Zou,
  %``Coupled-Channel Effects for the Bottomonium with Realistic Wave Functions,''
  Phys.\ Rev.\ D {\bf 94}, no. 3, 034021 (2016).      	
  
\bibitem{Jaffe:1976ih} 
  R.~L.~Jaffe,
  %``Multi-Quark Hadrons. 2. Methods,''
  Phys.\ Rev.\ D {\bf 15}, 281 (1977).  
  
\bibitem{Barbour:1979qi} 
  I.~M.~Barbour and D.~K.~Ponting,
  %``Nonstrange Four Quark States,''
  Z.\ Phys.\ C {\bf 5}, 221 (1980);
  I.~M.~Barbour and J.~P.~Gilchrist,
  %``The $N \bar{N}$ and $\pi \pi$ Decay Modes of Baryonium,''
  Z.\ Phys.\ C {\bf 7}, 225 (1981)
  Erratum: [Z.\ Phys.\ C {\bf 8}, 282 (1981)].
  
\bibitem{Weinstein:1983gd}
  J.~D.~Weinstein and N.~Isgur,
  %``The q q anti-q anti-q System in a Potential Model,''
  Phys.\ Rev.\ D {\bf 27}, 588 (1983).	   
  
\bibitem{SilvestreBrac:1993ss} 
  B.~Silvestre-Brac and C.~Semay,
  %``Systematics of L = 0 q-2 anti-q-2 systems,''
  Z.\ Phys.\ C {\bf 57}, 273 (1993).  
  
\bibitem{Brink:1998as} 
  D.~M.~Brink and F.~Stancu,
  %``Tetraquarks with heavy flavors,''
  Phys.\ Rev.\ D {\bf 57}, 6778 (1998).  
  
\bibitem{Maiani:2004vq}
  L.~Maiani, F.~Piccinini, A.~D.~Polosa and V.~Riquer,
  %``Diquark-antidiquarks with hidden or open charm and the nature of X(3872),''
  Phys.\ Rev.\ D {\bf 71}, 014028 (2005).
  
\bibitem{Barnea:2006sd} 
  N.~Barnea, J.~Vijande and A.~Valcarce,
  %``Four-quark spectroscopy within the hyperspherical formalism,''
  Phys.\ Rev.\ D {\bf 73}, 054004 (2006).  
  
\bibitem{Santopinto:2006my} 
  E.~Santopinto and G.~Galat\`a,
  %``Spectroscopy of tetraquark states,''
  Phys.\ Rev.\ C {\bf 75}, 045206 (2007).  
  
\bibitem{Anwar:2017toa} 
  M.~N.~Anwar, J.~Ferretti, F.~K.~Guo, E.~Santopinto and B.~S.~Zou,
  %``Spectroscopy and decays of the fully-heavy tetraquarks,''
  arXiv:1710.02540;
  M.~N.~Anwar, J.~Ferretti and E.~Santopinto,
  %``Spectroscopy of the hidden-charm $\boldsymbol{[qc][\bar q \bar c]}$ and $\boldsymbol{[sc][\bar s \bar c]}$ tetraquarks,''
  arXiv:1805.06276.  
  
\bibitem{Dubynskiy:2008mq} 
  S.~Dubynskiy and M.~B.~Voloshin,
  %``Hadro-Charmonium,''
  Phys.\ Lett.\ B {\bf 666}, 344 (2008).  
  
\bibitem{Guo:2008zg} 
  F.~K.~Guo, C.~Hanhart and U.~G.~Meissner,
  %``Evidence that the Y(4660) is a f(0)(980)psi-prime bound state,''
  Phys.\ Lett.\ B {\bf 665}, 26 (2008).  
  
\bibitem{Guo:2009id} 
  F.~K.~Guo, C.~Hanhart and U.~G.~Meissner,
  %``Implications of heavy quark spin symmetry on heavy meson hadronic molecules,''
  Phys.\ Rev.\ Lett.\  {\bf 102}, 242004 (2009).  
  
\bibitem{Voloshin:2013dpa} 
  M.~B.~Voloshin,
  %``$Z_c(3900)$ - what is inside?,''
  Phys.\ Rev.\ D {\bf 87}, no. 9, 091501 (2013).
  
\bibitem{Li:2013ssa} 
  X.~Li and M.~B.~Voloshin,
  %``$Y$(4260) and $Y$(4360) as mixed hadrocharmonium,''
  Mod.\ Phys.\ Lett.\ A {\bf 29}, no. 12, 1450060 (2014).  
  
\bibitem{Wang:2013kra} 
  Q.~Wang, M.~Cleven, F.~K.~Guo, C.~Hanhart, U.~G.~Meißner, X.~G.~Wu and Q.~Zhao,
  %``Y(4260): hadronic molecule versus hadro-charmonium interpretation,''
  Phys.\ Rev.\ D {\bf 89}, no. 3, 034001 (2014).  
  
\bibitem{Cleven:2015era} 
  M.~Cleven, F.~K.~Guo, C.~Hanhart, Q.~Wang and Q.~Zhao,
  %``Employing spin symmetry to disentangle different models for the XYZ states,''
  Phys.\ Rev.\ D {\bf 92}, no. 1, 014005 (2015).    
  
\bibitem{Brambilla:2015rqa} 
  N.~Brambilla, G.~Krein, J.~Tarrús Castellà and A.~Vairo,
  %``Long-range properties of $1S$ bottomonium states,''
  Phys.\ Rev.\ D {\bf 93}, no. 5, 054002 (2016).    
  
\bibitem{Alberti:2016dru} 
  M.~Alberti, G.~S.~Bali, S.~Collins, F.~Knechtli, G.~Moir and W.~Söldner,
  %``Hadroquarkonium from lattice QCD,''
  Phys.\ Rev.\ D {\bf 95}, no. 7, 074501 (2017).  
  
\bibitem{Panteleeva:2018ijz} 
  J.~Y.~Panteleeva, I.~A.~Perevalova, M.~V.~Polyakov and P.~Schweitzer,
  %``On tetraquarks with hidden charm and strangeness as phi-psi(2S) hadrocharmonium,''
  arXiv:1802.09029.                           
  
\bibitem{Guo:2014iya}
  F.~K.~Guo, C.~Hanhart, Q.~Wang and Q.~Zhao,
  %``Could the near-threshold $XYZ$ states be simply kinematic effects?,''
  Phys.\ Rev.\ D {\bf 91}, 051504 (2015).

\bibitem{Szczepaniak:2015eza}
  A.~P.~Szczepaniak,
  %``Triangle Singularities and XYZ Quarkonium Peaks,''
  Phys.\ Lett.\ B {\bf 747}, 410 (2015).

\bibitem{Liu:2015taa}
  X.~H.~Liu, M.~Oka and Q.~Zhao,
  %``Searching for observable effects induced by anomalous triangle singularities,''
  Phys.\ Lett.\ B {\bf 753}, 297 (2016).  
  
\bibitem{Seth:2012}
  K.~K.~Seth,
  Prog.\ Part.\ Nucl.\ Phys.\ {\bf 67}, 390 (2012).
	
\bibitem{Esposito:2014rxa} 
  A.~Esposito, A.~L.~Guerrieri, F.~Piccinini, A.~Pilloni and A.~D.~Polosa,
  %``Four-Quark Hadrons: an Updated Review,''
  Int.\ J.\ Mod.\ Phys.\ A {\bf 30}, 1530002 (2015). 		
  
\bibitem{Olsen:2017bmm} 
  S.~L.~Olsen, T.~Skwarnicki and D.~Zieminska,
  %``Nonstandard heavy mesons and baryons: Experimental evidence,''
  Rev.\ Mod.\ Phys.\  {\bf 90}, no. 1, 015003 (2018).   
  
\bibitem{Guo:2017jvc} 
  F.~K.~Guo, C.~Hanhart, U.~G.~Meißner, Q.~Wang, Q.~Zhao and B.~S.~Zou,
  %``Hadronic molecules,''
  Rev.\ Mod.\ Phys.\  {\bf 90}, no. 1, 015004 (2018).  
  
\bibitem{Gottfried:1977gp} 
  K.~Gottfried,
  %``Hadronic Transitions Between Quark - anti-Quark Bound States,''
  Phys.\ Rev.\ Lett.\  {\bf 40}, 598 (1978).
					
\bibitem{Voloshin:1978hc} 
  M.~B.~Voloshin,
  %``On Dynamics of Heavy Quarks in Nonperturbative QCD Vacuum,''
  Nucl.\ Phys.\ B {\bf 154}, 365 (1979).					
	
\bibitem{Peskin:1979va} 
  M.~E.~Peskin,
  %``Short Distance Analysis for Heavy Quark Systems. 1. Diagrammatics,''
  Nucl.\ Phys.\ B {\bf 156}, 365 (1979);
  G.~Bhanot and M.~E.~Peskin,
  %``Short Distance Analysis for Heavy Quark Systems. 2. Applications,''
  Nucl.\ Phys.\ B {\bf 156}, 391 (1979).  
  
\bibitem{Aubert:2005rm} 
  B.~Aubert {\it et al.} [BaBar Collaboration],
  %``Observation of a broad structure in the $\pi^+ \pi^- J/\psi$ mass spectrum around 4.26-GeV/c$^2$,''
  Phys.\ Rev.\ Lett.\  {\bf 95}, 142001 (2005).
  
\bibitem{Mizuk:2009da} 
  R.~Mizuk {\it et al.} [Belle Collaboration],
  %``Dalitz analysis of B ---> K pi+ psi-prime decays and the Z(4430)+,''
  Phys.\ Rev.\ D {\bf 80}, 031104 (2009).
  
\bibitem{Wang:2007ea} 
  X.~L.~Wang {\it et al.} [Belle Collaboration],
  %``Observation of Two Resonant Structures in e+e- to pi+ pi- psi(2S) via Initial State Radiation at Belle,''
  Phys.\ Rev.\ Lett.\  {\bf 99}, 142002 (2007).
  
\bibitem{Aubert:2007zz} 
  B.~Aubert {\it et al.} [BaBar Collaboration],
  %``Evidence of a broad structure at an invariant mass of 4.32- $GeV/c^{2}$ in the reaction $e^{+} e^{-} \to \pi^{+} \pi^{-} \psi_{2S}$ measured at BaBar,''
  Phys.\ Rev.\ Lett.\  {\bf 98}, 212001 (2007).    
  
\bibitem{Ablikim:2013wzq} 
  M.~Ablikim {\it et al.} [BESIII Collaboration],
  %``Observation of a Charged Charmoniumlike Structure $Z_c$(4020) and Search for the $Z_c$(3900) in $e^+e^- \to ?^+?^-h_c$,''
  Phys.\ Rev.\ Lett.\  {\bf 111}, no. 24, 242001 (2013).  
  
\bibitem{Eides:2015dtr} 
  M.~I.~Eides, V.~Y.~Petrov and M.~V.~Polyakov,
  %``Narrow Nucleon-$\psi(2S)$ Bound State and LHCb Pentaquarks,''
  Phys.\ Rev.\ D {\bf 93}, no. 5, 054039 (2016);
  I.~A.~Perevalova, M.~V.~Polyakov and P.~Schweitzer,
  %``On LHCb pentaquarks as a baryon-$\psi$(2S) bound state: prediction of isospin-$\frac3{2}$ pentaquarks with hidden charm,''
  Phys.\ Rev.\ D {\bf 94}, no. 5, 054024 (2016).    
  
\bibitem{Brodsky:1989jd} 
  S.~J.~Brodsky, I.~A.~Schmidt and G.~F.~de Teramond,
  %``Nuclear Bound Quarkonium,''
  Phys.\ Rev.\ Lett.\  {\bf 64}, 1011 (1990).  
  
\bibitem{Luke:1992tm} 
  M.~E.~Luke, A.~V.~Manohar and M.~J.~Savage,
  %``A QCD Calculation of the interaction of quarkonium with nuclei,''
  Phys.\ Lett.\ B {\bf 288}, 355 (1992).  
  
\bibitem{Kaidalov:1992hd} 
  A.~B.~Kaidalov and P.~E.~Volkovitsky,
  %``Heavy quarkonia interactions with nucleons and nuclei,''
  Phys.\ Rev.\ Lett.\  {\bf 69}, 3155 (1992).            			  
	
\bibitem{Voloshin:2007dx}   
  M.~B.~Voloshin,
  %``Charmonium,''
  Prog.\ Part.\ Nucl.\ Phys.\  {\bf 61}, 455 (2008).	
	
\bibitem{Sibirtsev:2005ex} 
  A.~Sibirtsev and M.~B.~Voloshin,
  %``The Interaction of slow J/psi and psi' with nucleons,''
  Phys.\ Rev.\ D {\bf 71}, 076005 (2005).	 
  
\bibitem{Guo:2006ya} 
  F.~K.~Guo, P.~N.~Shen and H.~C.~Chiang,
  %``Chromo-polarizability and pi pi final state interaction,''
  Phys.\ Rev.\ D {\bf 74}, 014011 (2006).  
  
\bibitem{Yan:1980}    
  T.M.~Yan,
  Phys.\ Rev.\ D {\bf 22}, 1652 (1980).  
  
\bibitem{Yan:1981}
  Y.~P.~Kuang and T.M.~Yan,
  Phys.\ Rev.\ D {\bf 24}, 2874 (1981).    
  
\bibitem{Li:1986}
  G.~Z.~Li and Y.~P.~Kuang,
  Commun.\ Theor.\ Phys. {\bf 5}, 79 (1986). 
  
\bibitem{Yan:1988}
  Y.~P.~Kuang. S.~F.~Tuan and T.M.~Yan,
  Phys.\ Rev.\ D {\bf 37}, 1210 (1988).   
  
\bibitem{Zhou:1991}
  H.~Y.~Zhou and Y.~P.~Kuang,
  Phys.\ Rev.\ D {\bf 44}, 756 (1991).	 
  
\bibitem{Voloshin:1980zf} 
  M.~B.~Voloshin and V.~I.~Zakharov,
  %``Measuring QCD Anomalies in Hadronic Transitions Between Onium States,''
  Phys.\ Rev.\ Lett.\  {\bf 45}, 688 (1980).

\bibitem{Moxhay:1987ch} 
  P.~Moxhay,
  %``Hadronic Transitions of d Wave Quarkonium,''
  Phys.\ Rev.\ D {\bf 37}, 2557 (1988).  
  
\bibitem{Tye}
  S.-H.~H.~Tye,
  Phys.\ Rev.\ D {\bf 13}, 3416 (1976);
  R.~C.~Giles and S.-H.~H.~Tye,
  Phys.\ Rev.\ Lett.\ {\bf 37}, 1175 (1976);
  Phys.\ Rev.\ D {\bf 16}, 1079 (1977).	

\bibitem{Isgur-Paton}  
  N.~Isgur and J.~Paton,
  Phys.\ Rev.\ D {\bf 31}, 2910 (1985).   
  
\bibitem{Richard}
  J.-M. Richard, 
  Phys. Rep. {\bf 212}, 1 (1992). 

\bibitem{Basdevant}
  S. Boukraa and J.-L. Basdevant,
  J. Math. Phys. {\bf 30}, 1060 (1989).   
  
\bibitem{Feynman-Lectures}
  R.~P.~Feynman, R.~B.~Leighton and M. L. Sands,
  {\it The Feynman Lectures on Physics}, 
  Addison-Wesley Pub. Co. (1963-1965).      	 
  
\bibitem{delAmoSanchez:2010jr}   
  K.~Abe {\it et al.} [Belle Collaboration],
  %``Observation of a near-threshold omega J/psi mass enhancement in exclusive B ---> K omega J/psi decays,''
  Phys.\ Rev.\ Lett.\  {\bf 94}, 182002 (2005);
  B.~Aubert {\it et al.} [BaBar Collaboration],
  %``Observation of Y(3940) $\to J/\psi \omega$ in $B \to J/\psi \omega K$ at BABAR,''
  Phys.\ Rev.\ Lett.\  {\bf 101}, 082001 (2008);
  P.~del Amo Sanchez {\it et al.} [BaBar Collaboration],
  %``Evidence for the decay X(3872) ---> J/ psi omega,''
  Phys.\ Rev.\ D {\bf 82}, 011101 (2010).
  
\bibitem{Uehara:2009tx} 
  S.~Uehara {\it et al.} [Belle Collaboration],
  %``Observation of a charmonium-like enhancement in the gamma gamma ---> omega J/psi process,''
  Phys.\ Rev.\ Lett.\  {\bf 104}, 092001 (2010);
  J.~P.~Lees {\it et al.} [BaBar Collaboration],
  %``Study of $X(3915) \to J/\psi \omega$ in two-photon collisions,''
  Phys.\ Rev.\ D {\bf 86}, 072002 (2012). 
  
\bibitem{Abe:2007jna} 
  K.~Abe {\it et al.} [Belle Collaboration],
  %``Observation of a new charmonium state in double charmonium production in e+ e- annihilation at s**(1/2) ~ 10.6-GeV,''
  Phys.\ Rev.\ Lett.\  {\bf 98}, 082001 (2007).  
  
\bibitem{Abe:2007sya} 
  P.~Pakhlov {\it et al.} [Belle Collaboration],
  %``Production of New Charmoniumlike States in e+ e- --> J/psi D(*) anti-D(*) at s**(1/2) ~ 10. GeV,''
  Phys.\ Rev.\ Lett.\  {\bf 100}, 202001 (2008).  	
  
\bibitem{Lees:2012cn} 
  J.~P.~Lees {\it et al.} [BaBar Collaboration],
  %``Study of the reaction $e^{+}e^{-} \to J/\psi\pi^{+}\pi^{-}$ via initial-state radiation at BaBar,''
  Phys.\ Rev.\ D {\bf 86}, 051102 (2012).  
  
\bibitem{He:2006kg} 
  Q.~He {\it et al.} [CLEO Collaboration],
  %``Confirmation of the Y(4260) resonance production in ISR,''
  Phys.\ Rev.\ D {\bf 74}, 091104 (2006).    
  
\bibitem{Yuan:2007sj} 
  C.~Z.~Yuan {\it et al.} [Belle Collaboration],
  %``Measurement of e+ e- ---> pi+ pi- J/psi cross-section via initial state radiation at Belle,''
  Phys.\ Rev.\ Lett.\  {\bf 99}, 182004 (2007). 
  
\bibitem{Lees:2012pv} 
  J.~P.~Lees {\it et al.} [BaBar Collaboration],
  %``Study of the reaction $e^{+}e^{-}\to \psi(2S)\pi^{-}\pi^{-}$ via initial-state radiation at BaBar,''
  Phys.\ Rev.\ D {\bf 89}, no. 11, 111103 (2014).  
  
\bibitem{Wang:2014hta} 
  X.~L.~Wang {\it et al.} [Belle Collaboration],
  %``Measurement of $e^+e^- \to \pi^+\pi^-\psi(2S)$ via Initial State Radiation at Belle,''
  Phys.\ Rev.\ D {\bf 91}, 112007 (2015).       
  
\bibitem{Ablikim:2016qzw} 
  M.~Ablikim {\it et al.} [BESIII Collaboration],
  %``Precise measurement of the $e^+e^-\to \pi^+\pi^-J/\psi$ cross section at center-of-mass energies from 3.77 to 4.60 GeV,''
  Phys.\ Rev.\ Lett.\  {\bf 118}, no. 9, 092001 (2017).    	  		
			
\end{thebibliography}
\end{document}